\documentclass[conference]{IEEEtran}
\usepackage{amsfonts}
\IEEEoverridecommandlockouts

\ifCLASSINFOpdf
\else
\fi
\usepackage{epsfig}
\usepackage{graphicx}
\usepackage{psfig}
\usepackage{epsf}
\usepackage[cmex10]{amsmath}
\usepackage{booktabs}
\usepackage{fancyhdr}
\usepackage{slashbox}
\usepackage{caption2}   

\newtheorem{rem}{Remark} 
\begin{document}
\title{Enhancing PHY Security of MISO NOMA SWIPT Systems With a Practical Non-Linear EH Model}
\author{Fuhui Zhou$^{*\pounds} $, Zheng Chu$^\dag$, Yongpeng Wu$^\S$, Naofal Al-Dhahir$^\ddag$, Pei Xiao$^\dag$\\
$^*$Nanchang University, Nanchang, China, $^\pounds$Xidian University, Xi'an, China, $^\dag$University of Surrey, Guildford, U.K \\
 $^\S $Shanghai Jiao Tong University, China, $^\ddag$University
of Texas at Dallas, Dallas, USA\\

Email: \emph{\{zhoufuhui@ieee.org,  \{andrew.chuzheng7, yongpeng.wu2016\}@gmail.com, aldhahir@utdallas.edu\}}
\thanks{The research was supported by the National Natural Science Foundation of China (61701214, 61701301, 61661028), the Foundation of The State Key Laboratory of ISN (ISN19-08), the Young Natural Science Foundation of Jiangxi Province (20171BAB212002), and the Postdoctoral Science Foundation (2017M610400, 2017KY04, 2017RC17). The work of N. Al-Dhahir was made possible by NPRP grant $\#$ NPRP 8-627-2-260  from the Qatar National Research Fund. }}
\maketitle

\begin{abstract}
Non-orthogonal multiple-access (NOMA) and simultaneous wireless information and power transfer (SWIPT) are promising techniques to improve spectral efficiency and energy efficiency. However, the security of NOMA SWIPT systems has not received much attention in the literature. In this paper, an artificial noise-aided beamforming design problem is studied to enhance the security of a multiple-input single-output NOMA SWIPT system where a practical non-linear energy harvesting model is adopted. The problem is non-convex and challenging to solve. Two algorithms are proposed to tackle this problem based on semidefinite relaxation (SDR) and successive convex approximation. Simulation results show that a performance gain can be obtained by using NOMA compared to the conventional orthogonal multiple access. It is also shown that the performance of the algorithm using a cost function is better than the algorithm using SDR at the cost of a higher computation complexity.
\end{abstract}
\begin{IEEEkeywords}
Non-orthogonal multiple access, physical-layer security, resource allocation, non-linear energy harvesting model.
\end{IEEEkeywords}
\IEEEpeerreviewmaketitle
\section{Introduction}
\IEEEPARstart{T}{HE} escalating high data rate requirements and the unprecedented increase of mobile devices call for the fifth generation wireless communication systems to address challenging issues, such as spectrum scarcity and massive connectivity \cite{J. G. Andrews}, \cite{F. Zhou}. Non-orthogonal multiple access (NOMA) is a promising technique to address these challenges \cite{Z. Ding}. NOMA can provide high spectral efficiency (SE) and simultaneously serve multiple users. Unlike orthogonal multiple access (OMA), successive interference cancellation (SIC) techniques are required to reduce the mutual interference among different users due to the exploitation of non-orthogonal resources \cite{Z. Wei}. NOMA has received increasing attention since it can achieve a significant performance gain in terms of SE and energy efficiency (EE) compared with OMA \cite{Z. Ding1}-\cite{Y. Zhang}.

Recently, another promising technique called simultaneous wireless information and power transfer (SWIPT) has been proposed to prolong the operational lifetime of energy-limited devices, such as  energy-limited sensors \cite{M. Ku}. Unlike the conventional energy harvesting (EH) techniques, such as solar and wind power, SWIPT can provide stable and controllable power for wireless applications. Particularly, radio frequency (RF) signals not only are identified as sources for energy harvesting, but also they carry the information required for transmission. Thus, in NOMA systems with energy-limited devices, it is of great importance to investigate the efficient integration of SWIPT with NOMA that can simultaneously improve SE and prolong the lifetime of the system \cite{Y. Liu}, \cite{P. Diamantoulakis}. However, due to the broadcasting nature of NOMA and the dual function of RF signals, malicious energy harvesting receivers (EHRs) may exist and intercept the confidential transmitted information signals \cite{F. Zhou2}-\cite{E. Boshkovska3}. Thus, it is vital to improve the security of NOMA SWIPT systems.

Physical-layer security (PHY) has attracted great interests recently since it does not require extra resources for the secret key \cite{F. Zhou2}-\cite{E. Boshkovska3}. It exploits the physical layer characteristics of the wireless channels to achieve secure communication. However, it was shown that the secrecy rate of wireless communication systems is limited by the channel state information \cite{Y. Zou1}. In order to improve the secrecy rate, multiple antennas, cooperative relaying, jamming and artificial noise (AN)-aided techniques have been investigated \cite{Y. Liu1}. It was shown that these techniques are efficient to improve the secrecy rate. Although many investigations have been conducted to improve the security of OMA SWIPT systems, to the authors' best knowledge, this paper is the first work focusing on improving the security of NOMA SWIPT systems. In this paper, an AN-aided beamforming design problem is studied in multiple-input single-output (MISO) NOMA SWIPT systems. The related works are summarized next.

In OMA SWIPT systems, the existing works fall into two categories according to the EH model, namely, the linear EH model \cite{F. Zhou2}, \cite{D. W. K. Ng2}, \cite{Z. Chu} and the non-linear EH model \cite{E. Boshkovska3}. AN-aided beamforming design problems were studied in MISO cognitive radios \cite{F. Zhou2} and multiple-input multiple-output (MIMO) systems \cite{D. W. K. Ng2}, \cite{Z. Chu}, respectively, and an ideal linear EH model was applied. In this EH model, the output power is linearly increased with the input power. Obviously, this model is unrealistical due to the practical nonlinear end-to-end power conversation circuit \cite{M. Ku}. Recently, in \cite{E. Boshkovska3}, a practical non-linear EH model was proposed and the optimal beamforming scheme was designed in secure MISO systems based on physical-layer security. Since NOMA is very different from OMA, the beamforming schemes proposed in \cite{F. Zhou2}, \cite{E. Boshkovska3}, \cite{D. W. K. Ng2} and \cite{Z. Chu} are unsuitable for securing NOMA systems based on physical-layer security. To improve the security of NOMA systems, optimal resource allocation schemes were designed in single-input single-output (SISO) \cite{Y. Zhang1}, MISO \cite{Y. Li} and MIMO NOMA systems \cite{M. Tian}, respectively. In \cite{Y. Zhang1}, the optimal power allocation strategy was proposed to maximize the secrecy rate while the optimal beamforming schemes were designed for maximizing the secrecy rate in \cite{Y. Li} and \cite{M. Tian}. It was shown that the secrecy rate achieved by using NOMA is higher than that obtained with OMA and that the performance can be significantly improved by using multiple antennas techniques.

However, the secure resource allocation schemes proposed in \cite{Y. Zhang1}-\cite{M. Tian} for NOMA systems are not adaptable to NOMA SWIPT systems. In a NOMA SWIPT system, the EH requirement needs to be considered.  Moreover, challenges remain to be addressed in this system, especially when a practical non-linear EH model is adopted. Unlike the linear EH model, the harvested power under the non-linear EH model is in a complex non-linear form involving exponential functions \cite{E. Boshkovska}, \cite{E. Boshkovska3}. Thus, it is an interesting yet challenging problem to study resource allocation in NOMA SWIPT systems.

In this paper, in order to improve security and energy conversion efficiency of NOMA  SWIPT systems, AN-aided and multiple antennas techniques are exploited. A power minimization problem is formulated in MISO NOMA  SWIPT systems based on a practical non-linear EH model. The transmission beamforming and AN-aided covariance matrix are jointly optimized to minimize the transmission power of the base station (BST). It is challenging to solve this non-convex optimization problem. Two suboptimal schemes based on successive convex approximation (SCA) are proposed to solve this problem. Simulation results show that the performance achieved by using NOMA is better than that obtained by using OMA. It is also shown that the proposed cost function method can achieve a performance gain over the proposed semidefinite relaxation (SDR)-based method at the cost of a higher implementation complexity.

The remainder of this paper is organized as follows. Section II presents the system model. The AN-aided beamforming design problem is formulated in Section III. Section IV presents simulation results. The paper concludes with Section V.

\emph{Notations:} Boldface capital letters and boldface lower case letters denote matrices and vectors, respectively. $\mathbf{I}$ denotes the identity matrix. The Hermitian (conjugate) transpose, trace, and rank of a matrix \textbf{A} are represented respectively by $\mathbf{A^H}$, Tr$\left(\mathbf{A}\right)$ and Rank$\left(\mathbf{A}\right)$. The conjugate transpose of a vector $\mathbf{x}$ is denoted by $\mathbf{x}^\dag$. $\mathbf{C}^{M\times N}$ represents a $M$-by-$N$ dimensional complex matrix set. $\mathbf{A}\succeq \mathbf{0} \left(\mathbf{A}\succ \mathbf{0}\right)$ represents that $\mathbf{A}$ is a Hermitian positive semidefinite (definite) matrix. $\mathbb{H}^N$ and $\mathbb{H}_+^{N}$ denote a $N$-by-$N$ dimensional Hermitian matrix set and a Hermitian positive semidefinite matrix set, respectively. ${\left\|  \cdot  \right\|}$ denotes the Euclidean norm of a vector. The absolute value of a complex scalar is denoted by ${\left| \cdot \right|}$. $\mathbf{x} \sim {\cal C}{\cal N}\left( {\mathbf{u},\mathbf{\Sigma } }\right)$ means that $\mathbf{x}$ is a random vector and follows a complex Gaussian distribution with mean $\mathbf{u}$ and covariance matrix $\mathbf{\Sigma }$. $\mathbb{E}[ \cdot ]$ denotes the expectation operator. $\mathbb{R}_{+}$ denotes the set of all nonnegative real numbers. $\min \left( {a,b} \right)$ denotes the minimum value between $a$ and $b$. ${\lambda _{\max }}\left( {{\mathbf{A}}}\right) $ is the maximum eigenvalue of $\mathbf{A}$.
\section{System Model}
\begin{figure}[!t]
\centering
\includegraphics[width=2.8 in]{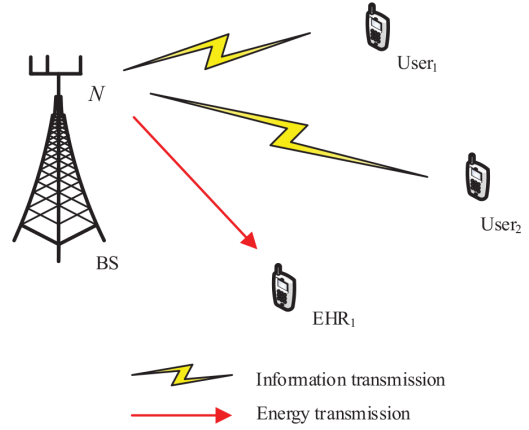}
\caption{The system model.} \label{fig.1}
\end{figure}
A downlink  MISO NOMA SWIPT system is depicted in Fig. 1, where a BST transmits information to two users (e.g., user 1 and user 2) and simultaneously transfers energy to an energy harvesting receiver (EHR) by using the SWIPT technique. The BST is equipped  with $N$ antennas. Both the users and the EHR are equipped with a single antenna. Due to the broadcast nature of NOMA and the dual function (information and energy transmission) of RF signals, the EHR may eavesdrop and intercept the information transmitted by the BST. All the channels involved are assumed to be frequency-non-selective fading channels. Channel state information (CSI) is assumed to be perfect \cite{Y. Zhang1}-\cite{M. Tian}.  Note that the CSI can be obtained by estimating it at the EHR and then sending it back to the BST through a feedback link (assumed error-free in this simplified model) \cite{Y. Zhang1}-\cite{M. Tian}. The performance achieved under these  assumptions can be used as upper-bound analysis and provides some meaningful insights for the design of MISO SWIPT NOMA.

In order to improve the security of the considered system, an AN-aided scheme is adopted. In this case, the signals received at the $j$th user, where $j  \in \left\{ {1,2} \right\}$, and the EHR denoted by ${y_j}$ and ${y_e}$, respectively, are given as
\begin{subequations}
\begin{align}\label{27}\
&{y_j} = \mathbf{h}_j^\dag \sum\limits_{i = 1}^2 {\left( {{\mathbf{w}_i}{s_i} + \mathbf{v}} \right)}  + {n_j},\\
&{y_e} = {\mathbf{g}^\dag }\sum\limits_{i = 1}^2 {\left( {{\mathbf{w}_i}{s_i} + \mathbf{v}} \right)}  + {n_e},
\end{align}
\end{subequations}
where $\mathbf{h}_j\in {\mathbf{C}^{{N} \times 1}}$ and $\mathbf{g}\in {\mathbf{C}^{{N} \times 1}}$ denote the channel impulse response vectors between the BST and the $j$th user, and the EHR, respectively. In $\left(1\right)$, $s_i\in {\mathbf{C}^{{1} \times 1}}$ and $\mathbf{w}_i\in {\mathbf{C}^{{N} \times 1}}$ are the confidential information-bearing signal for the $i$th user and the corresponding beamforming vector, respectively. It is assumed that $\mathbb{E}[ {{{\left| s \right|}^2}} ] = 1$; $\mathbf{v}\in {\mathbf{C}^{{N} \times 1}}$ is the AN vector generated by the BST in order to improve the secrecy rate of users and the harvested energy at the EHR. It is assumed that $\mathbf{v} \sim {\cal C}{\cal N}\left( {0,\mathbf{\Sigma} } \right)$ and $\mathbf{\Sigma}$ is the AN covariance matrix to be designed. In $\left(1\right)$, ${n_j} \sim {\cal C}{\cal N}\left( {0,\sigma _j^2} \right)$ and ${n_{e}} \sim {\cal C}{\cal N}\left( {0,\sigma _e^2} \right)$  represent the complex Gaussian noise at the $j$th user and the EHR, respectively.

It is assumed that $\left\| {{\mathbf{h}_1}} \right\| \ge \left\| {{\mathbf{h}_2}} \right\|$. According to the NOMA protocol, SIC is performed at User$_1$. Similar to \cite{Y. Zhang1}, it is assumed that User$_2$'s message is decoded before the EHR decodes User$_1$'s message. This action overestimates the EHR's interception capability.  Thus, according to \cite{Y. Zhang1}-\cite{M. Tian}, the secrecy rates of the $j$th user, denoted by $R_j$,  are given as
\begin{subequations}
\begin{align}\label{27}\
&{R_1} = {\log _2}\left( {1 + \frac{{\mathbf{h}_1^\dag {\mathbf{w}_1}\mathbf{w}_1^\dag {\mathbf{h}_1}}}{{\mathbf{h}_1^\dag \mathbf{\Sigma} {\mathbf{h}_1}+\sigma _1^2}}} \right) - {\log _2}\left( {1 + \frac{{{\mathbf{g}^\dag }{\mathbf{w}_1}\mathbf{w}_1^\dag \mathbf{g}}}{{{\mathbf{g}^\dag }\mathbf{\Sigma} \mathbf{g} + \sigma _e^2}}} \right)\\
&{R_2} = R - {\log _2}\left( {1 + \frac{{{\mathbf{g}^\dag }{\mathbf{w}_2}\mathbf{w}_2^\dag \mathbf{g}}}{{{\mathbf{g}^\dag }\left( {{\mathbf{w}_1}\mathbf{w}_1^\dag  + \mathbf{\Sigma} } \right)\mathbf{g }+ \sigma _e^2}}} \right)\\
&R = \min \left\{ {{{\log }_2}\left( {1 + \frac{{\mathbf{h}_2^\dag {\mathbf{w}_2}\mathbf{w}_2^\dag {\mathbf{h}_2}}}{{\mathbf{h}_2^\dag \left( {{\mathbf{w}_1}\mathbf{w}_1^\dag  + \mathbf{\Sigma} } \right){\mathbf{h}_2} + \sigma _2^2}}} \right),} \right.\\
&\ \ \ \ \ \ \left. {{{\log }_2}\left( {1 + \frac{{\mathbf{h}_1^\dag {\mathbf{w}_2}\mathbf{w}_2^\dag {\mathbf{h}_1}}}{{\mathbf{h}_1^\dag \left( {{\mathbf{w}_1}\mathbf{w}_1^\dag  + \mathbf{\Sigma} } \right){\mathbf{h}_1} + \sigma _1^2}}} \right)} \right\}.
\end{align}
\end{subequations}
According to the practical non-linear EH model \cite{E. Boshkovska}, \cite{E. Boshkovska3}, the harvested power at EHR denoted by ${\Phi _{e}}$, is given as
\begin{subequations}
\begin{align}\label{27}\
&{\Phi _{e}} =  {\frac{{{\psi _{e}} - P_{e}^{\max }{\Psi _{e}}}}{{1 - {\Psi _{e}}}}} ,\\
&{\psi _{e}} = \frac{{P_{e}^{\max }}}{1 + {\exp\left[{ - {a}\left( {{\Gamma _{e}} - {b}} \right)}\right]}},\\
&{\Psi _{e}} = \frac{1}{1 + \exp\left({{a}{b}}\right)},\\
&{\Gamma _e} = {\mathbf{g}^\dag }\left( {\sum\limits_{i = 1}^2 {{\mathbf{w}_i}\mathbf{w}_i^\dag }  + \mathbf{\Sigma} } \right)\mathbf{g}.
\end{align}
\end{subequations}
$P_{e}^{\max }$ is the maximum harvested power of the EHR when the EHR circuit is saturated; $a$ and $b$ are the parameters that reflect the circuit specifications, such as the capacitance, the resistance, and diode turn-on voltage \cite{E. Boshkovska}.

\section{AN-aided Beamforming Design Problem in MISO NOMA SWIPT Systems}
\subsection{Problem Formulation}
The power minimization problem subject to the secrecy rate constraint and the EH constraint denoted by $\text{P}_{{1}}$, is given as
\begin{subequations}
\begin{align}\label{27}\
\text{P}_{{1}}: \ \ \ \ \ &{\mathop {\min }\limits_{\mathbf{w}_i,\mathbf{\Sigma} } }\ {\text{Tr}\left( {\sum\limits_{i = 1}^2 {{\mathbf{w}_i}\mathbf{w}_i^\dag }  + \mathbf{\Sigma} } \right)}\\
&\text{s.t.}\ \ \  C1:{R_i} \ge {\gamma _i}, i  \in \left\{ {1,2} \right\},\\
&\ \ \ \ \ \ \  C2:{\Phi _e} \ge {\Upsilon _e},\\
&\ \ \ \ \ \ \  C3:\mathbf{\Sigma} \succeq0.
\end{align}
\end{subequations}
${\gamma _i}$ is the minimum required secrecy rate of the $i$th user and $\Upsilon _e$ is the minimum EH requirement of the EHR. The constraint $C1$ can guarantee that the secrecy rate of the $i$th user is no less than ${\gamma _i}$ and the constraint $C2$ is the EH constraint of the EHR. Due to the constraints $C1$ and $C2$, $\text{P}_{{1}}$ is non-convex and difficult to solve. To overcome this problem, two schemes based on SCA are proposed.
\subsection{Two Suboptimal Solutions}
\subsubsection{The SDR-Based Scheme}
Let ${\mathbf{W}_i} = {\mathbf{w}_i}\mathbf{w}_i^\dag$, where $i  \in \left\{ {1,2} \right\}$, ${\mathbf{H}_i} = {\mathbf{h}_i}\mathbf{h}_i^\dag$ and ${\mathbf{G}} = {\mathbf{G}}\mathbf{G}^\dag$. By using SDR, $\text{P}_{{1}}$ can be relaxed to $\text{P}_{{2}}$, given as
\begin{subequations}
\begin{align}\label{27}\
&\text{P}_{{2}}:  \ {\mathop {\min }\limits_{\mathbf{W}_i,\mathbf{\Sigma} } }\ {\text{Tr}\left( {\sum\limits_{i = 1}^2 {{\mathbf{W}_i} }  + \mathbf{\Sigma} } \right)} \\ \notag
&\text{s.t.}\ \ \ \\
& C4: \frac{{\left( {\text{Tr}\left( {\mathbf{\Sigma} {\mathbf{H}_1}} \right) + \sigma _1^2} \right)\left\{ {\text{Tr}\left[ {\left( {{\mathbf{W}_1} + \mathbf{\Sigma }} \right)\mathbf{G}} \right] + \sigma _e^2} \right\}}}{{\left\{ {\text{Tr}\left[ {\left( {{\mathbf{W}_1} + \mathbf{\Sigma} } \right){\mathbf{H}_1}} \right] + \sigma _1^2} \right\}\left[ {\text{Tr}\left( {\mathbf{\Sigma} \mathbf{G}} \right) + \sigma _e^2} \right]}} \le {2^{-{\gamma _1}}},\\ \notag
& C5:\\ \notag
&\frac{{\left\{ {\text{Tr}\left[ {\left( {{\mathbf{W}_1} + \mathbf{\Sigma} } \right){\mathbf{H}_i}} \right] + \sigma _i^2} \right\}\left\{ {\text{Tr}\left[ {\left( {\sum\limits_{i = 1}^2 {{\mathbf{W}_i}}  + \mathbf{\Sigma} } \right)\mathbf{G}} \right] + \sigma _e^2} \right\}}}{{\left\{ {\text{Tr}\left[ {\left( {\sum\limits_{i = 1}^2 {{\mathbf{W}_i}}  + \mathbf{\Sigma} } \right){\mathbf{H}_i}} \right] + \sigma _i^2} \right\}\left\{ {\text{Tr}\left[ {\left( {{\mathbf{W}_1} + \mathbf{\Sigma} } \right)\mathbf{G}} \right] + \sigma _e^2} \right\}}},\\ \notag \\
& \le {2^{ - {\gamma _2}}}, i \in \left\{ {1,2} \right\},\\
& C6:\text{Tr}\left[ {\left( {\sum\limits_{i = 1}^2 {{\mathbf{W}_i}}  + \mathbf{\Sigma} } \right)\mathbf{G}} \right] \ge c, \\
& C7:\mathbf{\Sigma} \succeq0, \mathbf{W}_i \succeq0,
\end{align}
\end{subequations}
where $c$ is given as
\begin{align}\label{27}\
c = {b} - \frac{{\ln \left\{ {\frac{{P_e^{\max }}}{{{\Upsilon _e}\left( {1 - {\Psi _e}} \right) + P_e^{\max }{\Psi _e}}} - 1} \right\}}}{a}.
\end{align}
$\text{P}_{{2}}$ is still non-convex due to constraints $C4$ and $C5$. In order to make $\text{P}_{{2}}$ tractable, exponential auxiliary variables are applied to convert constraints $C4$ and $C5$ into the following equivalent forms \cite{Chu3}.  Constraint $C4$ can be equivalently expressed as
\begin{subequations}
\begin{align}\label{27}\
&\exp \left( {{x_1} + {x_2} - {y_1} - {y_2}} \right) \le {2^{ - {\gamma _1}}},\\
&{\text{Tr}\left( {\mathbf{\Sigma} {\mathbf{H}_1}} \right) + \sigma _1^2} \le \exp \left( {{x_1}} \right),\\
&{\text{Tr}\left[ {\left( {{\mathbf{W}_1} + \mathbf{\Sigma }} \right)\mathbf{G}} \right] + \sigma _e^2} \le \exp \left( {{x_2}} \right),\\
&{\text{Tr}\left[ {\left( {{\mathbf{W}_1} + \mathbf{\Sigma} } \right){\mathbf{H}_1}} \right] + \sigma _1^2} \ge \exp \left( {{y_1}} \right),\\
&{\text{Tr}\left( {\mathbf{\Sigma} \mathbf{G}} \right) + \sigma _e^2} \ge \exp \left( {{y_2}} \right),
\end{align}
\end{subequations}
where ${x_1}$, ${x_2}$, ${y_1}$  and ${y_2}$ are auxiliary variables. The constraint given by $\left(7\right)$ is still non-convex due to the constraints given by $\left(7\rm{b}\right)$ and $\left(7\rm{c}\right)$. To address this issue, a Taylor series expansion is applied. Thus, the constraints given by $\left(7\rm{b}\right)$ and $\left(7\rm{c}\right)$ can be approximated as
\begin{subequations}
\begin{align}\label{27}\
&{\text{Tr}\left( {\mathbf{\Sigma} {\mathbf{H}_1}} \right) + \sigma _1^2} \le \exp \left( {{{\widetilde x}_1}} \right)\left( {{x_1} - {{\widetilde x}_1} + 1} \right),\\
&{\text{Tr}\left[ {\left( {{\mathbf{W}_1} + \mathbf{\Sigma }} \right)\mathbf{G}} \right] + \sigma _e^2} \le \exp \left( {{{\widetilde x}_2}} \right)\left( {{x_2} - {{\widetilde x}_2} + 1} \right),
\end{align}
\end{subequations}
where ${\widetilde x}_1$ and ${\widetilde x}_2$ are approximate values, and they are equal to ${x_1}$ and ${x_2}$, respectively, when the constraints are tight. Similar to $C4$, $C5$ can be approximated as
\begin{subequations}
\begin{align}\label{27}\
&\exp \left( {{z_i} + {x_3} - {q_i} - {y_3}} \right) \le {2^{ - {\gamma _2}}},\\
& {\text{Tr}\left[ {\left( {{\mathbf{W}_1} + \mathbf{\Sigma} } \right){\mathbf{H}_i}} \right] + \sigma _i^2} \le \exp \left( {{{\widetilde z}_i}} \right)\left( {{z_i} - \widetilde {{z_i}} + 1} \right),\\
&{\text{Tr}\left[ {\left( {\sum\limits_{i = 1}^2 {{\mathbf{W}_i}}  + \mathbf{\Sigma} } \right)\mathbf{G}} \right] + \sigma _e^2} \le \exp \left( {\widetilde {{x}}_3} \right)\left( {{x_3} - \widetilde {{x}}_3 + 1} \right), \\
&{\text{Tr}\left[ {\left( {\sum\limits_{i = 1}^2 {{\mathbf{W}_i}}  + \mathbf{\Sigma} } \right){\mathbf{H}_i}} \right] + \sigma _i^2}\ge \exp \left( {{q_i}} \right),\\
&{\text{Tr}\left[ {\left( {{\mathbf{W}_1} + \mathbf{\Sigma} } \right)\mathbf{G}} \right] + \sigma _e^2}  \ge \exp \left( {{y_3}} \right),
\end{align}
\end{subequations}
where $z_i$, ${x_3}$, ${q_i}$  and ${y_3}$ are auxiliary variables. ${\widetilde z}_i$ and ${\widetilde x}_3$ are approximate values, and they are equal to ${z_i}$ and ${x_3}$, respectively, when the corresponding constraints are tight. Thus, $\text{P}_{{2}}$ can be approximated as $\text{P}_{{3}}$, given below
\begin{subequations}
\begin{align}\label{27}\
&\text{P}_{{3}}:  \ {\mathop {\min }\limits_{\mathbf{W}_i,\mathbf{\Sigma}, x_k, y_k,z_i, q_i } }\ {\text{Tr}\left( {\sum\limits_{i = 1}^2 {{\mathbf{W}_i} }  + \mathbf{\Sigma} } \right)} ,\\
&\text{s.t.}\ \ \ \left(7\rm{a}\right), \left(7\rm{d}\right), \left(7\rm{e}\right), \left(8\right), \left(9\right), C6, C7,
\end{align}
\end{subequations}
where $i\in \left\{ {1,2} \right\}$ and $k\in\left\{ {1,2,3} \right\}$. It can bee seen that $\text{P}_{{3}}$ is convex and can be efficiently solved by using the software package \texttt{CVX} \cite{F. Zhou2}. Based on the solution of $\text{P}_{{3}}$, an iterative algorithm using SCA can be developed to solve $\text{P}_{{1}}$, denoted by Algorithm 1. The details of Algorithm 1 are seen in Table 1, where $P_{opt}^n$ is the total transmit power at the $n$th iteration.

\begin{rem}
Algorithm 1 cannot guarantee that the solution of $\text{P}_{{1}}$, denoted by $\mathbf{W}_i$, is a rank-one matrix. If the solution $\mathbf{W}_i$ is rank-one, the optimal robust secure beamforming vector for $\text{P}_{{1}}$ can be  obtained using the eigenvalue decomposition. If the solution $\mathbf{W}_i$ is not rank-one, the suboptimal robust secure beamforming vector can be obtained by using the well-known Gaussian randomization procedure \cite{F. Zhou2}.
\end{rem}
\begin{table}[htbp]
\begin{center}
\caption{The SDR-based algorithm}
\begin{tabular}{lcl}
\\\toprule
$\textbf{Algorithm 1}$: The SDR-based algorithm for $\text{P}_{{1}}$\\ \midrule
\  1: \textbf{Setting:}\\
\ \  \ $\gamma _i$, $i\in \{1,2\}$, $\Upsilon _e$ and the tolerance error $\xi$; \\
\  2: \textbf{Initialization:}\\
\ \  \ The iterative number $n=1$, ${\widetilde x}_i^n$, ${\widetilde y}_i^n$, ${\widetilde q}_i^n$, ${\widetilde z}_i^n$, and $P_{opt}^n$; \\
\  3: \textbf{Repeat:}\\
 \   \ \ \ solve $\text{P}_{\textbf{3}}$ by using \texttt{CVX} for the given approximate values; \\
\ \ \ \ \ obtain ${\widetilde x}_i^{\left(n+1\right)}$, ${\widetilde y}_i^{\left(n+1\right)}$, ${\widetilde q}_i^{\left(n+1\right)}$, ${\widetilde z}_i^{\left(n+1\right)}$;\\
 \ \ \ \ \ if rank $\mathbf{W}_i$=1 \\
  \ \ \ \ \ \ Obtain optimal $\mathbf{W}_i$ and $\mathbf{\Sigma}$;  \\
 \ \ \ \ \ else \\
 \ \ \ \ \ \ Obtain suboptimal $\mathbf{W}_i$ and $\mathbf{\Sigma}$;   \\
 \ \ \ \ \ end \\
 \ \ \ \ \ update the iterative number $n=n+1$;  \\
 \ \ \ \ \ calculate the total transmit power $P_{opt}^n$;  \\
\ \ \ \ \ if $\left|P_{opt}^n-P_{opt}^{\left(n-1\right)}\right|\leq \xi$ \\
\ \ \ \ \  \  break;\\
\ \ \ \ \ end;\\
\  4: \textbf{Obtain resource allocation:}\\
 \ \ \ \ \ \ $\mathbf{w}_i$ and $\mathbf{\Sigma}$. \\
\bottomrule
\end{tabular}
\end{center}
\end{table}
\subsubsection{The Cost Function-based Scheme}

Since $\mathbf{W}_i$ is positive semi-definite, the rank of $\mathbf{W}_i$ equals $1$ when its maximum eigenvalue is equal to its trace, namely, $\text{Rank}\left(\mathbf{W}_i\right)=\lambda_{\max}\left(\mathbf{W}_i\right)$; otherwise, $\text{Rank}\left(\mathbf{W}_i\right)>\lambda_{\max}\left(\mathbf{W}_i\right)$. Thus, the rank-one constraint is equivalent to the constraint $\sum\limits_{i = 1}^2 {\left[ {\text{Tr}\left( {{\mathbf{W}_i}} \right) - {\lambda _{\max }}\left( {{\mathbf{W}_i}} \right)} \right]}  \le 0$. With this insight, we can see that the smaller $\sum\limits_{i = 1}^2 {\left[ {\text{Tr}\left( {{\mathbf{W}_i}} \right) - {\lambda _{\max }}\left( {{\mathbf{W}_i}} \right)} \right]} $ is, the more likely that the rank one constraints can be satisfied. By adopting a cost function, $\text{P}_{{3}}$ is reformulated into $\text{P}_{{4}}$  as follows
\begin{subequations}
\begin{align}\label{27}\ \notag
&\text{P}_{{4}}:  {\mathop {\min }\limits_{\mathbf{W}_i,\mathbf{\Sigma},x_i, y_i,z_i, q_i } }\ \text{Tr}\left( {\sum\limits_{i = 1}^2 {{\mathbf{W}_i} }  + \mathbf{\Sigma} } \right)+ \\
&\ \ \ \ \ \ \ \ \ \ \ \ \ \ \ \alpha\sum\limits_{i = 1}^2 {\left[ {\text{Tr}\left( {{\mathbf{W}_i}} \right) - {\lambda _{\max }}\left( {{\mathbf{W}_i}} \right)} \right]}  \\
&\ \ \ \ \ \ \ \text{s.t.}\ \ \ \left(10\rm{b}\right),
\end{align}
\end{subequations}
where $\alpha>0$ is a cost factor. The minimum value $\sum\limits_{i = 1}^2 {\left[ {\text{Tr}\left( {{\mathbf{W}_i}} \right) - {\lambda _{\max }}\left( {{\mathbf{W}_i}} \right)} \right]} $ can be obtained using a large  $\alpha$. Since ${\lambda _{\max }}\left( {{\mathbf{W}_i}}\right) $ is convex, $\text{P}_{{4}}$ is non-convex. The following lemma is applied to solve the non-convex problem.

\emph{\text{Lemma 1 }} \cite{S. P. Boyd}: Let ${\lambda _{\max }}\left( {{\mathbf{X}}}\right) $ and  ${\lambda _{\max }}\left( {{\mathbf{Y}}}\right) $ denote the maximum eigenvalues of $\mathbf{X}$ and $\mathbf{Y}$, respectively. If $\mathbf{X}$ and $\mathbf{Y}$ are positive semi-definite, ${\lambda _{\max }}\left( \mathbf{X} \right) - {\lambda _{\max }}\left( \mathbf{Y} \right) \ge \mathbf{y}_{\max }^\dag \left( {\mathbf{X} - \mathbf{Y}} \right){\mathbf{y}_{\max }}$, where ${\mathbf{y}_{\max }}$ is the eigenvector corresponding to the maximum eigenvalue of ${\mathbf{Y}}$.

Using Lemma 1, $\text{P}_{{4}}$ can be approximated as $\text{P}_{{5}}$, given below
\begin{subequations}
\begin{align}\label{27}\
&\text{P}_{{5}}:  {\mathop {\min }\limits_{\mathbf{W}_i,\mathbf{\Sigma},x_k, y_k,z_i, q_i } }\ {f\left( {\mathbf{W}_i^{n + 1}} \right)} \\ \notag
&\ \ \ \ \ \ \ \text{s.t.}\ \ \ \left(11\rm{b}\right),
\end{align}
\end{subequations}
where $n$ is the iteration index and ${f\left( {\mathbf{W}_i^{n + 1}} \right)}$ is given as
\begin{align}\label{27}\ \notag
f\left( {\mathbf{W}_i^{n + 1}} \right) = &\text{Tr}\left( {\sum\limits_{i = 1}^2 {\mathbf{W}_i^{n + 1}}  + {\mathbf{\Sigma} }} \right) + \\ \notag
&\alpha \sum\limits_{i = 1}^2 \left[ \text{Tr}\left( {\mathbf{W}_i^{n + 1}} \right) - {\lambda _{\max }}\left( {\mathbf{W}_i^n} \right)\right. \\
 &\ \ \ \ \ \ \ \ \ \ \ \left. - {{\left( {\mathbf{w}_i^n} \right)}^\dag }\left( {\mathbf{W}_i^{n + 1} - \mathbf{W}_i^n} \right)\mathbf{w}_i^n \right],
\end{align}
where $\mathbf{w}_i^n$ is the eigenvector corresponding to the maximum eigenvalue of ${\mathbf{W}_i^n}$. By solving $\text{P}_{{5}}$, an iterative algorithm, denoted by Algorithm 2, can be applied to solve $\text{P}_{{4}}$. The details of Algorithm 2 are presented in Table 2.
\begin{table}[htbp]
\begin{center}
\caption{The cost function-based algorithm}
\begin{tabular}{lcl}
\\\toprule
$\textbf{Algorithm 2}$: The cost function-based algorithm for $\text{P}_{{4}}$\\ \midrule
\  1: \textbf{Setting:}\\
\ \  \ $\gamma _i$, $i\in \{1,2\}$, $\Upsilon _e$, the tolerance error $\xi$ and $\alpha>0$; \\
\  2: \textbf{Initialization:}\\
\ \  \ The iterative number $n=1$, ${\widetilde x}_i^n$, ${\widetilde y}_i^n$, ${\widetilde q}_i^n$, ${\widetilde z}_i^n$, and $\mathbf{W}_i^{n }$; \\
\  3: \textbf{Repeat:}\\
 \   \ \ \ solve $\text{P}_{\textbf{5}}$ by using \texttt{CVX} for the given approximate values; \\
\ \ \ \ \ obtain ${\widetilde x}_i^{\left(n+1\right)}$, ${\widetilde y}_i^{\left(n+1\right)}$, ${\widetilde q}_i^{\left(n+1\right)}$, ${\widetilde z}_i^{\left(n+1\right)}$ and $\mathbf{W}_i^{\left(n+1\right)}$;\\
 \ \ \ \ \ if $\mathbf{W}_i^{\left(n+1\right)}$=$\mathbf{W}_i^{\left(n\right)}$;  \\
  \ \ \ \ \ set $\alpha=2\alpha$;\\
 \ \ \ \ \ end if  \\
 \ \ \ \ \ update the iterative number $n=n+1$;  \\
 \ \ \ \ \ calculate the total transmit power $P_{opt}^n$;  \\
\ \ \ \ \ if $\text{Tr}\left( {\mathbf{W}_i^{n + 1}} \right) - {\lambda _{\max }}\left( {\mathbf{W}_i^n} \right)\leq \xi$ \\
\ \ \ \ \  \  break;\\
\ \ \ \ \ end;\\
\  4: \textbf{Obtain resource allocation:}\\
 \ \ \ \ \ \ $\mathbf{w}_i$ and $\mathbf{\Sigma}$. \\
\bottomrule
\end{tabular}
\end{center}
\end{table}

Finally, the computation complexities of Algorithm 1 and Algorithm 2 are analyzed as follows. It is assumed that the number of iterations for SCA of Algorithm 1 and Algorithm 2 are denoted by $T_1$ and $T_2$, respectively. $\text{P}_{{3}}$ and $\text{P}_{{5}}$ have 3 linear matrix inequality (LMI) constraints of size $N$ and 15 linear constraints. The number of decision variables is $2N+10$. Thus, according to \cite{Chu3}, the computation complexity of Algorithm $i$, $i  \in \left\{ {1,2} \right\}$, is given by ${\rm O}\left( {n{T_i}\sqrt {3N + 15} \left[ {2\left( {{N^3} + n{N^2}} \right) + 15 + {n^2}} \right]} \right)$, where $n = {\rm O}\left( {2N + 10} \right)$ and ${\rm O}\left( {\cdot} \right)$ is the big ${\rm O}$ notation \cite{Chu3}.
\section{Simulation Results}
In this section,  simulation results are presented to compare the performance obtained by using NOMA to that achieved by using OMA. The convergence behavior of the proposed algorithms is also evaluated by simulation. The simulation settings are based on those used  in \cite{E. Boshkovska} and \cite{E. Boshkovska3}. All channels are assumed to be Rayleigh flat fading. The number of channel realizations is $10^4$ (if it is not special specified). The detailed simulation parameters are given in Table III.

\begin{table}[htbp]
 \caption{\label{tab:test}Simulation Parameters}
 \begin{tabular}{l|c|c}
  \midrule
  \midrule
  Parameters & Notation & Typical Values  \\
  \midrule
  \midrule
 Numbers of antennas & $N$ & $10$ \\
 The maximum harvested power & $P_{e}^{\max }$ & $24$ mW \\
 Circuit parameter & $a$ & $1500$ \\
 Circuit parameter & $b$ & $0.0022$ \\
 The minimum EH & $\Upsilon _e$ & $1$ mW\\
 The minimum secrecy rate & $\gamma _1$ & $3$ bits/s/Hz \\
 The minimum secrecy rate & $\gamma _2$ & $1.5$ bits/s/Hz \\
 Variances of Noise at Users & $\sigma _i^2$ & $-120$ dBm\\
 Variances of Noise at the EHR  & $\sigma _e^2$& $-120$ dBm\\
 The tolerance error & $\xi$ & $10^{-4}$ \\
   \midrule
 & ${ \mathbf{h}}_{1}$ & ${\cal C}{\cal N}\left( {\mathbf{0},2\mathbf{I } }\right)$ \\
  Channel distribution& ${ \mathbf{h}}_{2}$   & ${\cal C}{\cal N}\left( {\mathbf{0},1\mathbf{I } }\right)$\\
  & ${ \mathbf{g}}$   & ${\cal C}{\cal N}\left( {\mathbf{0},1.5\mathbf{I } }\right)$\\
\midrule
\midrule
 \end{tabular}
\end{table}

\begin{figure}[!t]
\centering
\includegraphics[width=3.0 in]{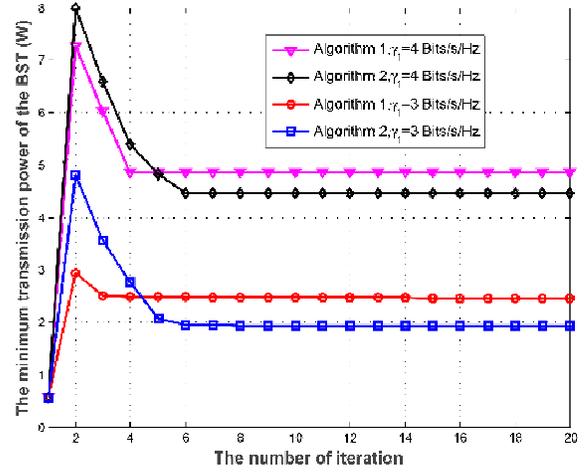}
\caption{The minimum transmission power of the BST versus the number of iterations under different algorithms, $\gamma _2=1.5$ Bits/s/Hz.} \label{fig.1}
\end{figure}
Fig. 2 shows the minimum transmission power of the BST versus the number of iterations required for Algorithms 1 and 2. The minimum required secrecy rate of User$_2$ is set to be 1.5 Bits/s/Hz. The minimum required secrecy rate of User$_1$ is set to be 3 Bits/s/Hz or 4 Bits/s/Hz. The minimum EH requirement of EHR is -30 dBW. The results are obtained by one channel realization. It can be seen  from Fig. 2 that only few iterations are required to achieve the minimum transmission power of the BST for both Algorithms 1 and 2. This indicates that Algorithms 1 and  2 are both computationally efficient. Moreover, as shown in Fig. 2, the number of iterations required for Algorithm 2 is larger than that for Algorithm 1. Since the computation complexity of Algorithms 1 and 2 is proportional to the number of iterations, we conclude that the computation complexity of Algorithm 2 is higher than that of Algorithm 1. This confirms that there is a tradeoff between the minimum transmission power of the BST and the computation complexity of the designed algorithm. It is also seen that the minimum transmission power of the BST achieved by using Algorithm 1 is larger than that obtained by using Algorithm 2. This demonstrates that Algorithm 2 provides a performance gain compared with Algorithm 1 in terms of the  minimum required transmission power. The reason is that Algorithm 2 can obtain rank-one solutions and achieve the global optimal solution.

\begin{figure}[!t]
\centering
\includegraphics[width=3.0 in]{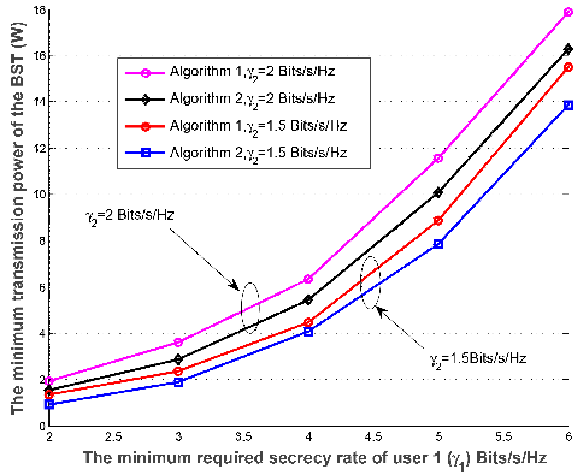}
\caption{The minimum transmission power of the BST versus the secrecy rate of the User$_1$ under different algorithms, $\gamma _2=1.5$ Bits/s/Hz or $\gamma _2=2$ Bits/s/Hz.} \label{fig.1}
\end{figure}
In order to further verify the performance advantage of Algorithm 2, Fig. 3 shows the minimum transmission power of the BST versus the secrecy rate of User$_1$ achieved by using Algorithms 1 and 2. $\gamma _2$ is set as 1.5 Bits/s/Hz or 2 Bits/s/Hz. As shown in Fig. 3, the performance of Algorithm 2 is better than that of Algorithm 1 in terms of the required transmission power. It is observed that the transmission power of the BST increases with the secrecy rate of User$_1$. The reason is that a higher power is required to satisfy the  the secrecy rate requirement of User$_1$.

\begin{figure}[!t]
\centering
\includegraphics[width=3.0 in]{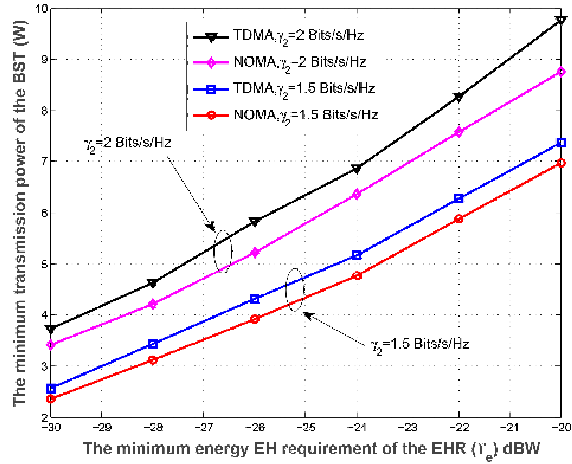}
\caption{The minimum transmission power of the BST versus the minimum EH requirement of the EHR achieved by using NOMA and TDMA, $\gamma _2=1.5$ Bits/s/Hz or $\gamma _2=2$ Bits/s/Hz.} \label{fig.1}
\end{figure}
Fig. 4 shows the minimum transmission power of the BST versus the minimum EH requirement of the EHR achieved by using NOMA and time division multiple access (TDMA). $\gamma _1$ is 3 Bits/s/Hz and $\gamma _2$ is set as 1.5 Bits/s/Hz or 2 Bits/s/Hz. For a better comparison between the performance achieved by using NOMA with that obtained by using TDMA, Algorithm 1 is applied. It is clearly seen that the minimum transmission power required by using NOMA is smaller than that required by using TDMA. The reason is that the spectrum efficiency of NOMA is higher than that of TDMA \cite{Z. Ding1}, \cite{Y. Zhang}, \cite{Y. Liu}. Thus, using NOMA, a smaller transmission power is required to satisfy the secrecy rate and energy harvesting constraints.
\section{Conclusion}
In this paper, in order to enhance the security of a MISO NOMA SWIPT system, an AN-aided beamforming design problem was studied under a practical non-linear EH model. The transmission beamforming and AN-aided covariance matrix were jointly optimized to minimize the transmit power of the BST while the secrecy rates of users and the EH requirement were satisfied. Two algorithms were proposed to solve this challenging non-convex problem. It was shown that the performance achieved by using NOMA is better than that obtained by using OMA. Simulation results also demonstrated that the algorithm based on the cost function is superior to the algorithm based on SDR at the cost of the computation complexity.


\begin{thebibliography}{20}
\bibitem{J. G. Andrews}
Z. Wei, \emph{et al.}, \lq\lq Optimal resource allocation for power-efficient MC-NOMA with imperfect channel state information?,\rq\rq \ \emph{IEEE Trans. Commun.}, vol. 65, no. 9, pp. 3944-3961, Sept. 2017.
\bibitem{F. Zhou}
F. Zhou, \emph{et al.}, \lq\lq Energy-efficient optimal power allocation for fading cognitive radio channels: ergodic capacity, outage capacity and minimum-rate capacity,\rq\rq \ \emph{IEEE Trans. Wireless Commun.}, vol. 15, no. 4, pp. 2741-2755, Apr. 2016.
\bibitem{Z. Ding}
L. Dai, \emph{et al.}, \lq\lq Joint pilot and payload power control for uplink MIMO-NOMA with MRC-SIC receivers,\rq\rq \ \emph{IEEE Commun. Lett.}, to appear, 2018.
\bibitem{Z. Wei}
Z. Wei, \emph{et al.}, \lq\lq Energy-efficient transmission design in Non-orthogonal multiple access,\rq\rq \ \emph{IEEE Trans. Veh. Technol.}, vol. 66, no. 3, pp. 2852-2857, Mar. 2017.
\bibitem{Z. Ding1}
B. Wang, \emph{et al.}, \lq\lq Spectrum and energy efficient beamspace MIMO-NOMA for millimeter-wave communications using lens antenna array,\rq\rq \ \emph{IEEE J. Sel. Areas Commun.}, to be published, 2017.
\bibitem{L. Liu}
L. Liu \emph{et al.}, \lq\lq Gaussian message passing iterative detection for MIMO-NOMA systems with massive users,\rq\rq \ in \emph{Proc. IEEE GLOBECOM 2016}, Washington, USA, Dec. 2016.
\bibitem{Y. Zhang}
Y. Zhang \emph{et al.}, \lq\lq Energy-efficient transmission design in Non-orthogonal multiple access,\rq\rq \ \emph{IEEE Trans. Veh. Technol.}, vol. 66, no. 3, pp. 2852-2857, Mar. 2017.
\bibitem{M. Ku}
Z. Zhu, \emph{et al.}, \lq\lq Outage constrained robust beamforming for secure broadcasting systems with energy harvesting,\rq\rq \ \emph{IEEE Trans. Wireless Commun.}, vol. 15, no. 11, pp. 7610-7620, Nov. 2016.
\bibitem{Y. Liu}
Y. Liu, \emph{et al.}, \lq\lq Cooperative nonorthogonal multiple access with simultaneous wireless information and power transfer\rq\rq \ \emph{IEEE J. Select. Areas Commun.}, vol. 34, no. 4, pp. 938-953, Apr. 2016.
\bibitem{P. Diamantoulakis}
Y. Xu, \emph{et al.}, \lq\lq Joint beamforming and power-splitting control in downlink cooperative SWIPT NOMA systems,\rq\rq \ \emph{IEEE Trans. Signal Proces.}, vol. 65, no. 18, pp. 4874-4886, Sept. 2017.
\bibitem{F. Zhou2}
F. Zhou, \emph{et al.}, \lq\lq Robust AN-Aided beamforming and power splitting design for secure MISO cognitive radio with SWIPT,\rq\rq \ \emph{IEEE Trans. Wireless Commun.}, vol. 16, no. 4, pp. 2450-2464, April 2017.
\bibitem{H. Wang}
H. Wang and X. Xia, \lq\lq Enhancing wireless secrecy via cooperation: signal design and optmization,\rq\rq \ \emph{IEEE Commun. Magz.}, vol. 53, no. 12, pp: 47-53, 2015.
\bibitem{E. Boshkovska}
E. Boshkovska, \emph{et al.}, \lq\lq Robust resource allocation for MIMO wireless powered communication networks based on a non-linear EH model,\rq\rq \ \emph{IEEE Trans. Commun.}, vol. 65, no. 5, pp. 1984-1999, May 2017.
\bibitem{E. Boshkovska3}
E. Boshkovska, \emph{et al.}, \lq\lq Secure SWIPT networks based on a non-linear energy harvesting model,\rq\rq \ in \emph{Proc. IEEE WCNC 2017},  San Francisco, CA, USA, 2017.
\bibitem{Y. Zou1}
X. Chen, \emph{et al.}, \lq\lq A survey on multiple-antenna techniques for physical layer security,\rq\rq \ \emph{IEEE Commun. Surveys Tuts.}, vol. 19, no. 2, pp. 1027-1053, Second Quarter, 2017.
\bibitem{Y. Liu1}
Y. Liu, \emph{et al.}, \lq\lq Phsical layer security for next generation wireless networks: Theories, tehcniques, and challenges,\rq\rq \ \emph{IEEE Commun. Surveys Tuts.}, vol. 19, no. 1, pp. 347-376, First Quarter, 2017.
\bibitem{D. W. K. Ng2}
D. W. K. Ng, \emph{et al.}, \lq\lq Robust beamforming for secure communication in systems with wireless information and power transfer,\rq\rq \ \emph{IEEE Trans. Wireless Commun.}, vol. 13, no. 8, pp. 4599-4615, Aug. 2014.
\bibitem{Z. Chu}
Z. Chu, \emph{et al.}, \lq\lq Secrecy rate optimizations for a MIMO secrecy channel with a cooperative jammer,\rq\rq \ \emph{IEEE Trans. Vehicular Technol.}, vol. 64, no. 5, pp. 1833-1847, May 2015.
\bibitem{Y. Zhang1}
Y. Zhang, \emph{et al.}, \lq\lq Secrecy sum rate maximization in non-orthogonal multiple access,\rq\rq \ \emph{IEEE Commun. Lett.}, vol. 20, no. 5, pp. 930-933, May 2016.
\bibitem{Y. Li}
Y. Li, \emph{et al.}, \lq\lq Secure beamforming in downlink MISO nonorthogonal multiple access systems,\rq\rq \ \emph{IEEE Trans.
Veh. Technol.}, to appear, 2017.
\bibitem{M. Tian}
M. Tian, \emph{et al.}, \lq\lq Secrecy sum rate optimization for downlink MIMO non-orthogonal multiple access systems,\rq\rq \ \emph{IEEE Signal Process. Lett.}, to be published, 2017.
\bibitem{S. P. Boyd}
S. P. Boyd and L. Vandenberghe, \emph{Convex Optimization}. Cambridge, U.K.: Cambridge Univ. Press, 2004.
\bibitem{Chu3}
Z. Chu, \emph{et al.}, \lq\lq Simulaneous wireless information power transfer for MISO secrecy channel,\rq\rq \ \emph{IEEE Trans. Vehicular Technol.}, vol. 65, no. 9, pp. 6913-6925, Sept. 2016.
\end{thebibliography}
\end{document}